# Reinterpretation of Hall effect in medium with hole

M. Oszwaldowski, P. Pieranski. T. Berus and J. Jankowski,

Instytut Fizyki, Politechnika Poznanska, ul. Nieszawska 13a, 60-965 Poznan, Poland.

Electronic address: maciej.oszwaldowski@put.poznan.pl

**Abstract**. We reinterpret the distribution of the Hall potential in the Hall bar-with-a-hole that has been found and interpreted by Mani and Klitzing [Appl. Phys. Lett., 64 1262 (1994)]. Our reinterpretation explains all the "paradoxes" without resorting to new theoretical conceptions.

## 1. Introduction

In their inspiring paper [1], G. R. Mani and K. von Klitzing demonstrated that an electronic system, named by them "Hall bar–with–a–hole" (HBWH), or "anti-Hall bar within a Hall bar" has very interesting and unexpected galvanomagnetic properties. Because of the hole, the system becomes a doubly connected one, with an inner area within the hole and an outer area outside the Hall bar. The authors pointed out that the understanding of the Hall effects in multiply connected systems is still not full and contains unresolved paradoxes. On the basis of the experimental results obtained for the HBWH, the authors suggested that the paradoxes could be explained through the identification of specific inversion symmetry in the system and assumption of a principle of superposition in the Hall effect. The existence of these two properties allows for a possibility of several independent Hall effects or resistances in the multiply connected system utilizing multiple boundary-injected currents. A particular outcome of this is the possibility of existing a classical Hall effect under null (net) current condition. The latter is rather difficult to accept. Nevertheless, in another paper [2], the authors presented an experimental evidence for the unexpected effect. They also suggested that it can be used for the construction of a new type of Hall sensor with strongly reduced and temperature independent voltage offset. Since the sensor works with full current compensation, it has strongly reduced heat dissipation. Consequently, the new Hall sensor can have an increased magnetic sensitivity by at least two orders of magnitude in comparison to the standard Hall sensor. In view of an important achievement, the novel sensor construction was patented [3].

Further experimental investigations [4-8] showed that the HBWH system is a very useful configuration for effective electron transport investigations, in particular for the Hall effect, both the 3D (classical) and the 2D (quantum) one. These investigations confirmed the validity of the inversion and the superposition operations [4]. The double-current studies revealed, however, that the Hall and the magnetoresistance electric potentials depend on the current injected into the sample in a different way. This was not understood.

More recently, J. Oswald and M. Oswald [9, 10] performed numerical calculations of the longitudinal (magnetoresistance) and the Hall voltages for the 2D and the 3D electron gas in the HBWH. They found an excellent agreement with the experimental results. They explained the agreement with the theoretical model in which the connectivity, the inversion symmetry and the superposition principle for the Hall and the longitudinal voltages played an essential role. Thus, they confirmed that one obtains simultaneously two independent voltages at the inner and the outer boundary of the HBWH system. Each Hall voltage depends exclusively on the current injected into the respective boundary. On the contrary, the longitudinal voltages are not boundary specific; they depend on the sum of both injected currents. Moreover, to interpret the quantum Hall effect (QHE) in the HBWH, they were lead to assume that the doubly connected HBWH structure is topologically equivalent to the doubly connected cylinder structure used by Laughlin to the explanation of the QHE. However, Laughlin theory predicts a bulk currier origin of the QHE, whereas Mani suggested the edge current origin for the QHE in the HBWH. Finally, the authors proposed a generalization of the Landauer-Buttiker formalism for the dissipative bulk transport that covers both the edge and the bulk effects. They also suggested that a theoretical approach to the QHE should take into account the sample topology.

The above short review of the results of the electron transport in the presence of magnetic field in the HBWH system shows that the studies resulted in a wealth of new ideas that are important for the 2D and the 3D electron transport in a multiply connected medium. In this context, the aim of this paper is to show that there is another, much simpler interpretation of the experimental facts observed in the HBWH that does not demand evoking these new ideas. In this proposed interpretation the notion of multiple-connectivity plays no role, because the Hall effects are associated with the standard, singly connected Hall bars.

## 2. Experimental results and their interpretation

For our comparison experiments, we used both the rectangular samples (the HBWH) studied experimentally by Mani as well as samples shaped as shown in Fig. 1. The sample onthe left hand side is an asymmetric HBWH having the left arm wider, and that on the right hand side is a symmetric circular sample, which also will be called HBWH. Both the samples were made of polycrystalline InSb thin films, about 2.4 μm thick. Their room temperature electron concentration was about $3\times10^{16}$ $cm^{-3}$ and the electron mobility was about 18,000 $cm^2/Vs$. The electric contact to the InSb films was made with silver paste. All measurements were performed at room temperature. We discuss here in detail the experimental results for the highly symmetric circular sample shown in Fig. 1. However, it should be pointed out that the results obtained for the rectangular and the circular samples were qualitatively the same. They were also in perfect qualitative agreement with the results obtained by Mani for his GaAs HBWH.

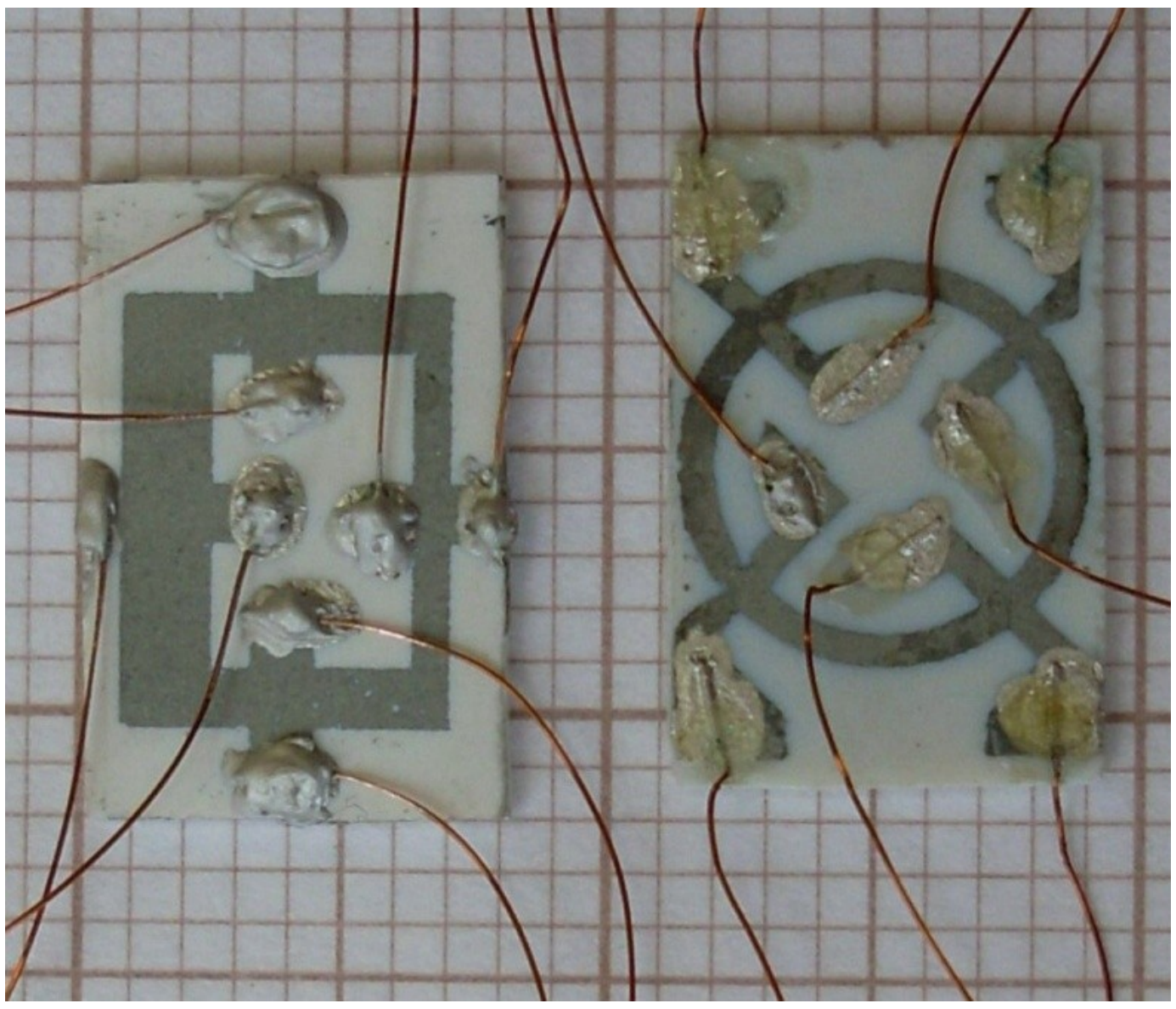

**Fig. 1** Photos taken from the experimentally investigated InSb layers with conducting wires attached. Left: asymmetric Hall bar with a hole. Right: symmetric Hall bar with a hole.

For a detailed description of the measurement results and their interpretation we begin with the single current injection. Figs. 2a-c show schematically the circular samples with the outer electric pads labelled by letters and the inner pads labelled by numbers. It is seen that the structure consists of four crosses; two of them are formed by the horizontal pads, *D-4* and *B-2*, and the others two are formed by the vertical pads, *A-1* and *C-3*. The current *I* is injected through the vertical pads. However, due to the full symmetry of the structure, the current could be equally injected through the horizontal pads. Our experiments show that when the current leads are attached to the external ends of the vertical pads, i. e. to the contacts *A* and *C*, then a Hall voltage appears across the external ends of the horizontal pads, i. e. at the contacts *B* and *D*. However, no Hall voltage appears across the inner contacts *2* and *4*. Similarly, if the current leads are connected to the internal contacts at the vertical pads, i. e. to the contacts *1* and *3*, then the same Hall voltage, but with opposite sign, appears across the internal contacts at the horizontal pad crosses, i.e. across contacts *2* and *4*, and no voltage appears across the external contacts *B* and *D*. These observations are in a full agreement with those of Mani.

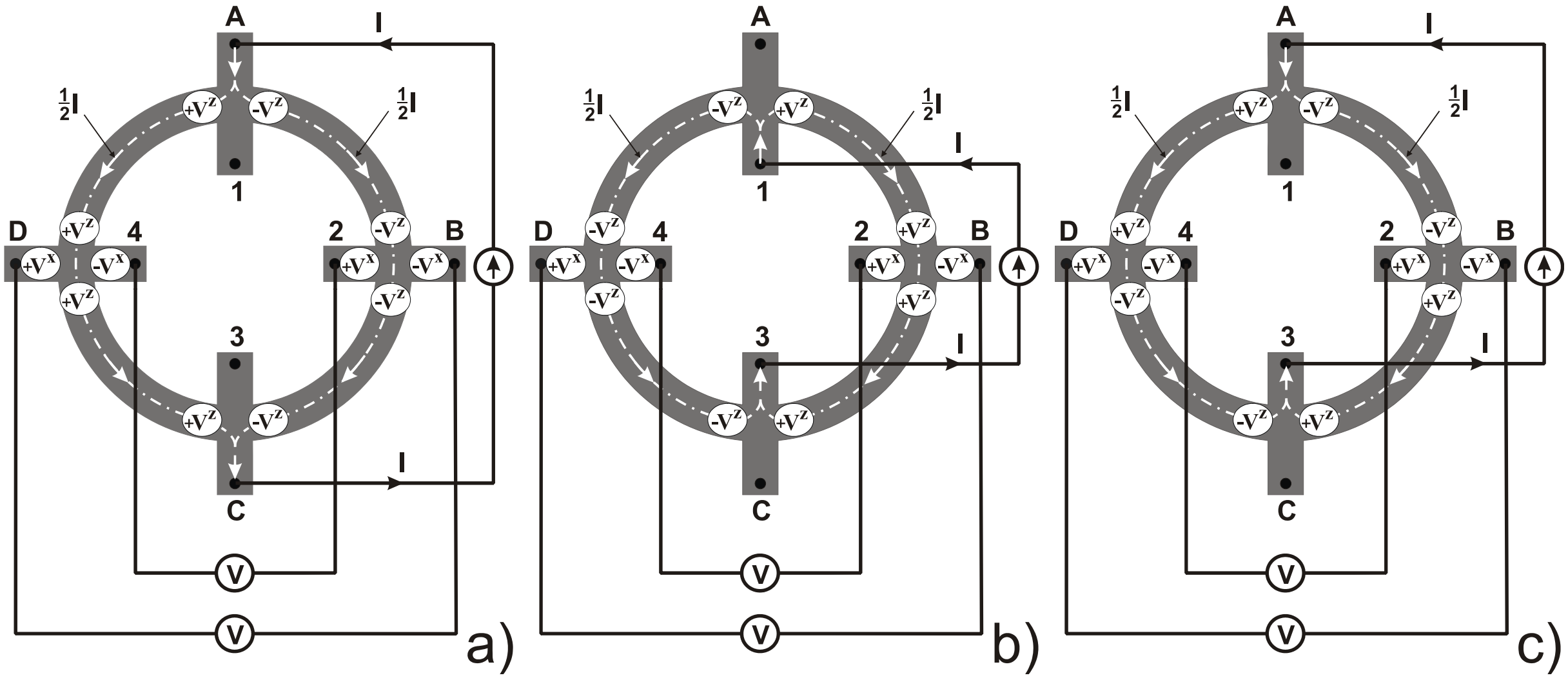

**Fig. 2** Hall effect measurement at various current injection configurations: (a) current *I* injected through external contacts *A* and *C*, (b) current injected through internal contacts *1* and *3*, and (c) current injected through external *A* and internal *3* contacts. Horizontal bar Hall crosses (left and right) generate Hall potentials $V^x$ and $-V^x$. Vertical bar Hall crosses (top and bottom) generate Hall potentials $V^y$ and $-V^y$.

In order to explain the rule that current injection through the external (internal) contacts generate a Hall voltage at the external (internal) contacts only, one should take into account the fact that in the system we have four Hall crosses, and all of them may be expected to contribute to the measured Hall voltages. Just in this point we differ in approach to the explanation of the Hall voltages first observed by Mani. In Refs [1, 2] only the two horizontal bar crosses were taken into account. Considering also the vertical bar crosses, one should observe that their wiring is untypical, and consequently their operation mode is also untypical. Comparison of the operation modes of the horizontal and the vertical bar Hall crosses is shown in Fig. 3.

In the analysis of measured Hall voltages we assume the following. The Hall potential is that generated in the system by a magnetic field. In the absence of magnetic field, the Hall potential is zero in whole the system. The magnetic field is directed in such a way that it deflects the current to the right, which means that the positive Hall potential $V^x$ is on the right hand side whereas the negative Hall potential $-V^x$ is on the left hand side of the current. We assume that the Hall potentials $V^x$ and $-V^x$ are observed, respectively, at contacts *D* and *2*, and *B* and *4* (Fig. 2) and they correspond to the current $I/2$.

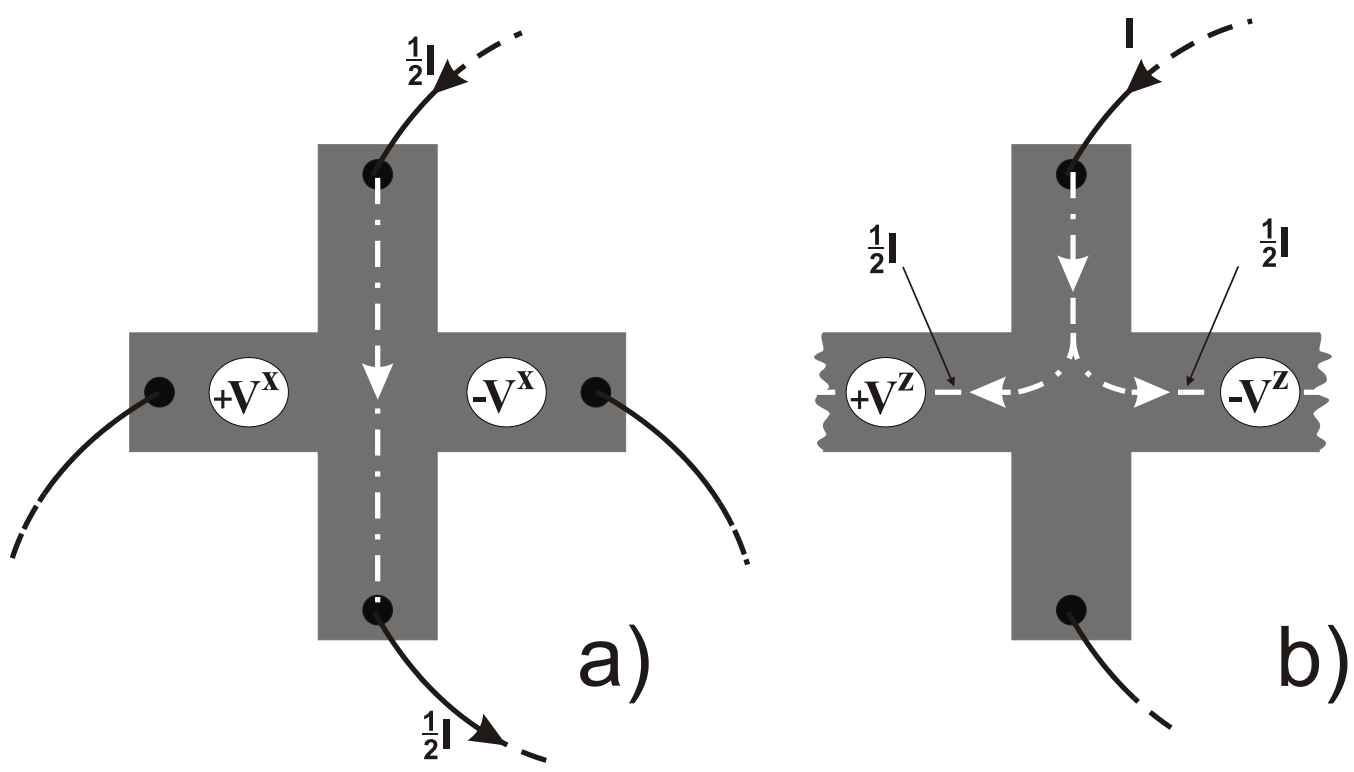

**Fig. 3** Two modes of operation of Hall crosses in Fig. 2 (a) standard mode, and (b) non-standard mode.

The operation mode shown in Fig. 3a is a standard operation mode for a Hall cross, in which a driving current $I/2$ generates a Hall voltage $U_H = 2V^x$. We assume that this voltage is a Hall voltage, because its measurement method fulfils the definition of the Hall voltage proposed by van der Pauw [11]. In contrast, the operation mode shown in Fig. 3b is different from the standard mode. In this mode, the current *I* generates Hall voltages $V^z$ and $-V^z$ , but the voltage $U^z = V^z - (-V^z)$ is not a Hall voltage, because its measuring method is not that given by the van der Pauw definition. The voltage $U^z$ can be determined in the HBWH configuration shown in Fig. 2, with the pairs of contacts *D* and *4*, and *B* and *2* short-circuited, in order to eliminate the voltages generated by the horizontal bar crosses. In this configuration $U^z$ is a voltage measured across the pairs of the short-circuited contacts. By doing this measurement, we determined $U^z$ and found that $V^z = V^x$ .

We can now apply this result to the analysis of the Hall voltages generated in the specific configurations of Fig. 2. Consider first the configurations shown in Fig. 2a and b. In both branches of each sample, the current $I/2$ flows in the same direction and generates on the right and on the left potentials $V^x$ and $-V^x$, respectively. Consequently, it generates the Hall voltages $U_H/2 = V^x - (-V^x)$ at contact pairs *D* and *4,* and *2* and *B*. Without any action of the vertical bar crosses those potentials can generate a Hall voltage $U_H/2$ on both the external contacts *B* and *D* and on the internal contacts *2* and *4*. Therefore, independent of whether the current is injected to the external *A* and *C* or to the internal *1* and *3* contacts, the Hall effect would be the same. However, the situation will be changed if the action of the vertical bar crosses is considered. If the current is injected at the external contacts *A* and *C*, it will increase the potential of the left branch by $V^z$ and will decrease the potential of the right branch by $-V^z$. Since $V^x = V^y$, the potentials at *2* and *4* reduce to zero. At the same time, the potential at *D* increases to $2V^x$ while the potential at *B* decreases to $-2V^x$. Thus, for the current injection at the external contacts *A* and *C*, the Hall voltage appearing at the external contacts *B* and *D* is $U_H = 4V^x$, and no voltage appears at the internal contacts *2* and *4*. This is also observed in the experiment.

In the case of the current injection at the internal contacts *1* and *3* (Fig. 2b), the potential of the left branch is decreased by $-V^z$ and that of the right branch potential is increased by $V^z$. Adding all potentials one finds that the potentials at the external contacts reduce to zero, the potential at *4* decreases to the value of $-2V^x$ and the potential at *2* increases to $2V^x$. Therefore, in agreement with the experiment, the current injection at internal contacts results in the appearance of the Hall voltage at the internal contacts only.

It may also be interesting to consider the case shown in Fig. 2c not discussed in the earlier literature. Here the current enters the sample through an external contact and leaves it through an internal contact. As may be seen in the figure, the action of both vertical bar Hall crosses leads to a mutual compensation at the horizontal bar crosses because in each branch one vertical bar cross generates potential $V^z$ and the other generates $-V^z$. As a result, the vertical bar Hall crosses do not contribute to the measured Hall voltage, leading to the same value $U_H = 2V^x$ on both the external (*D*, *B*) and the internal (*2*, *4*) contacts. This value is one half of that measured in the configurations (wirings) of Figs. 2a and b. The same value of $U_H$ can be observed in the case when the pairs of the contacts *A* and *1*, and *C* and *3* are short-circuited. The latter situation corresponds to a parallel connection of two independent Hall crosses formed by the left and the right branch of the HBWH of Fig. 2c.

Regarding the asymmetric HBWH having one branch wider than another (Fig. 1), we here also observed the rule that a current injected at the external (internal) contacts generates a Hall voltage only at the external (internal) contacts. Since the asymmetry in the branch width results in an asymmetry in the current distribution between the branches, current distribution does not affect the general Hall potential distribution. It is easy to show that the experimentally observed Hall voltages are possible only if the following relationships hold: $V_L^z = V_L^x$ and $V_P^z = V_P^x$, where superscripts *L* and *P* denote the left and the right branch of the sample.

Consider now the double injected current case. Mani [1] found experimentally that dual independent Hall effects may be realized simultaneously by injecting two independent currents into a HBWH, and that the Hall effects on each boundary (exterior or interior) reflect only the current injected via the same boundary. In these experiments, the current sources have to be galvanically separated. We repeated the experiments for the circular sample (Fig. 1) and found that the results are in a full agreement with those obtained by Mani. Injecting current $I_1$ into the current external contacts, we measured corresponding Hall voltage $U_H^1$ on the external voltage contacts. However, injecting current $I_2$ into the internal current contacts we measured corresponding Hall voltage $U_H^2$ on the external voltage contacts. The interpretation of this observation is straightforward. Since the current sources are independent, the Hall potentials in the double current condition must be described by a sum of the Hall voltages generated in single current conditions. These voltages were determined previously with the help of Figs 2a and b.

Interesting is the case when the currents are of the same magnitude but of opposite sign. This situation is shown in Fig. 4. Because both currents cancel along the ring, no net current flows there. The measured Hall voltages on the exterior *D* and *B* and the interior *2* and

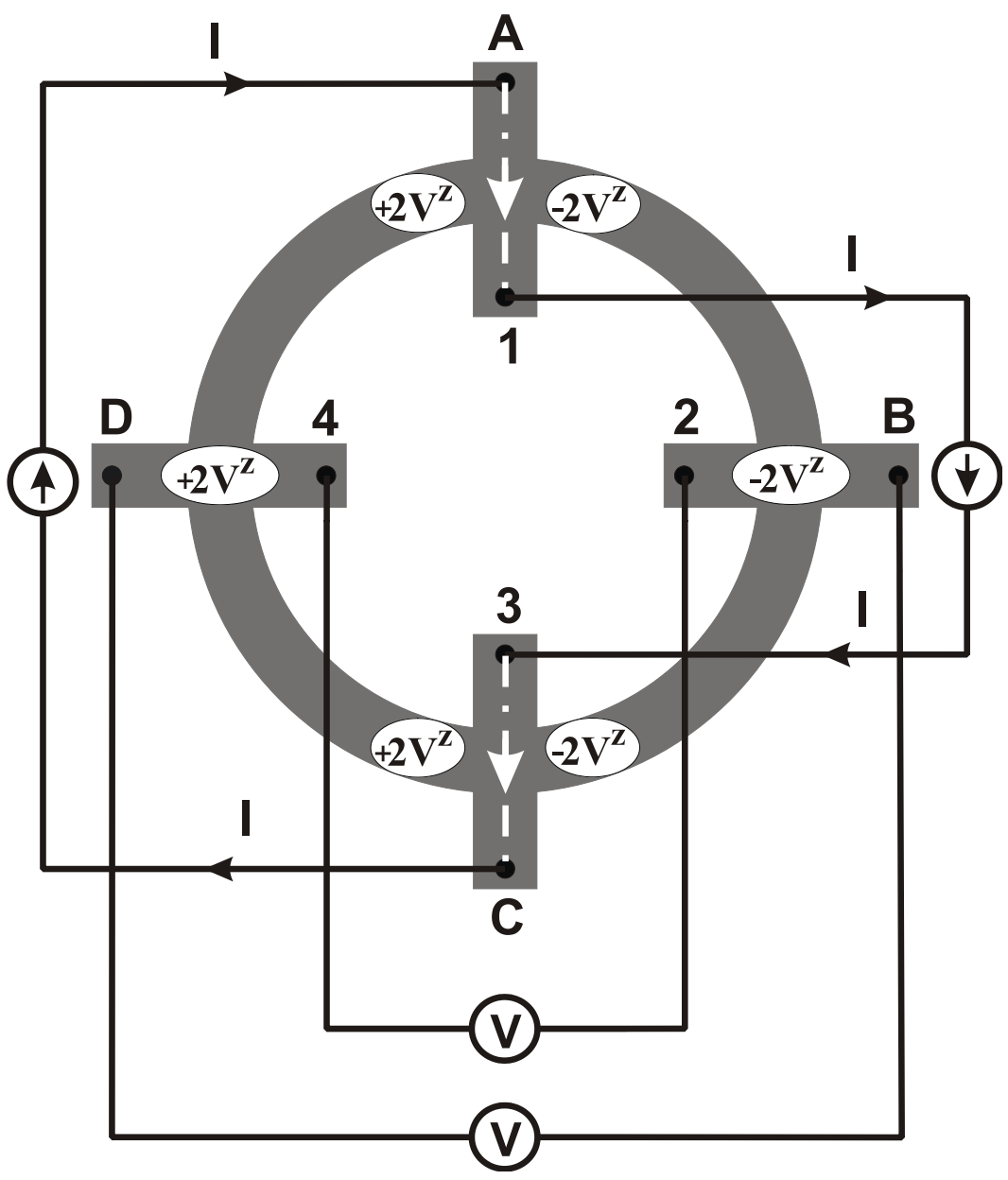

**Fig. 4** Hall effect under double-current conditions. Two equal currents *I* flow in opposite directions.

*4* electric contacts are due to the vertical bar Hall crosses that now operate in the standard mode of Fig. 3a. In this mode, the potential generated on the left branch is $2V^x$ and on the right branch is $-2V^x$. Thus, both on the internal and the external contacts the Hall voltage is $U_H = 4V^x$. Since the Hall voltages are generated by the current on the vertical bar Hall crosses, and not on the horizontal bar crosses where no current flows, the phrase “a Hall effect under null current conditions” [1] is not justified. Consequently, there is no reason to claim that the current cancellation makes the HBWH a favourable construction for Hall sensors because of greatly reduced heat dissipation [2]. Actually, the heat dissipation has to take place on the two Hall elements where the Hall voltage is in fact generated.

A main difficulty of the previous interpretation of the potential distribution in a HBWH was the observation that, in contrast to the Hall voltages, the magnetoresistance voltages are not sensitive to the origin of the current (interior or exterior) [1]. This disparity does not appear in the present interpretation. In order to explain the suggested disparity, one should notice that by definition the Hall voltages of a HBWH involve simultaneously potentials of both branches of the structure whereas the magnetoresistance voltages involve potentials of a single branch only. The Hall voltages are a sum of the Hall potentials generated simultaneously on the horizontal and the vertical bar crosses. The latter are origin sensitive, because any interchange between the internal and the external current injection contacts leads to a change in the current direction in the vertical bars and thereby to a change in the sign of the Hall potentials generated there. Therefore, the measured total Hall voltage must be origin sensitive. In contrast to this, the magnetoresistance voltages are by definition measured on a single branch. The vertical bar crosses add to a given branch a constant potential, changing sign with the change of the current origin. Since, the constant potential does not contribute to the magnetoresistance voltage, it is not origin sensitive.

## 3. Summary and conclusions

We have reinterpreted the potential distributions in a Hall bar-with-a-hole in crossed electric and magnetic fields. The reference point is the experimental data and their interpretation presented in Refs [1, 2] and further developed in Refs [3-10]. All Hall and magnetoresistance voltages predicted within the presented approach are in a full agreement with both the published as well as our own experimental data. In the reinterpretation we took into account two additional Hall crosses, which actively contribute to the potential distribution. These Hall crosses were not taken into account in the earlier interpretation probably because they work in a specific and untypical operation mode. In this specific mode the crosses generate a voltage, which is one-half of the Hall voltage generated in the standard operation mode.

The proposed reinterpretation is applicable also to a more general case of a medium with a hole having no specific geometrical symmetry. In contrast to the earlier interpretation, the proposed one is more complete as it also explains the problem of the difference in the sensitivity of the Hall and the magnetoresistance voltages to the current source location.

The previous difficulties in a clear understanding the Hall effects in the HBWH lead to the introduction of new notions and hypotheses such as specific inversion and the Hall effect superposition in a multiply connected medium. These new ideas appeared as an ingredient to the transport theory. From the point of view of the proposed reinterpretation, they remain needless. This is an important communication of the present paper.

**Acknowledgement**

This work was supported by a Politechnika Poznanska grant No BW–224/09.

**References**

[1] R. G. Mani and K. von Klitzing, Hall effect under null current conditions, Appl. Phys. Lett. 64 (1994), 1262-1264.

[2] R. G. Mani and K. von Klitzing, Temperature insensitive offset reduction in a Hall effect device, Appl. Phys. Lett. 64 (1994), 3121-3123.

[3] R. G. Mani and K. von Klitzing, Hall-Effect Device with Current and Hall-Voltage Connections, US Patent Nr 5,646,529, Jul. 8 1997.

[4] R. G. Mani, Experimental technique for realizing dual and multiple Hall effects in a single specimen, Europhys. Lett. 34 (1996), 139-144.

[5] R. G. Mani, Steady-state bulk current at high magnetic fields in Corbino-type GaAs/AlGaAs heterostructure devices, Europhys. Lett. 36 (1996), 203-208.

[6] R. G. Mani, Transport study of GaAs AlGaAs heterostructure and n-type GaAs devices in the “anti Hall bar within a Hall bar” configuration, J. Phys. Soc. Jpn. 65 (1996), 1751-1759.

[7] R. G. Mani, Dual Hall effects in inhomogeneous doubly connected GaAs/AlGaAsheterostructure devices, Appl. Phys. Lett. 70 (1997), 2879-2881.

[8] R. G. Mani, Dual ordinary, integral quantum, and fractional quantum Hall effects in partially gated doubly connected GaAs/$Al_xGa_{1-x}As$ heterostructure devices, Phys. Rev. B 55 (1997), 15838-15841.

[9] M. Oswald, J. Oswald and R. G. Mani, Voltage and current distribution in a doubly connected two-dimensional quantum Hall system, Phys. Rev. B 72 (2005), 035334-1 - 035334-7.

[10] J. Oswald and M. Oswald, Magnetotransport in a doubly connected two-dimensional quantum Hall system in the low magnetic field regime, Phys. Rev. B 74 **(**2006**),** 153315-1 - 153315-4.

[11] L. J. van der Pauw, A Method of Measuring Specific Resistivity and Hall Effect of Discs of Arbitrary Shape, Philips Res. Rep. 13 (1958), 1-9.